\documentclass[aps,amsmath, amssymb, prl, twocolumn, showpacs,superscriptaddress,groupedaddress]{revtex4-1}  
\pdfoutput=1
\usepackage{graphicx}  
\usepackage{dcolumn}   
\usepackage{bm}        
\usepackage{subfigure}
\usepackage{feynmp}

\newcommand{\gvec}[1]{\boldsymbol{#1}}
\newcommand{\rmd}{\textrm{d}}
\newcommand{\mut}{\gvec{\tilde{\mu}}}
\newcommand{\epsdt}{\tilde{\epsilon}_d}
\newcommand{\Ut}{\tilde{U}}
\newcommand{\Deltat}{\tilde{\Delta}}
\newcommand{\Sigmat}{\tilde{\Sigma}}

\newcommand\derivat[2]{\left.#1\right|_{#2}}

\newcommand{\sech}{\textrm{sech}}
\renewcommand{\Re}{\textrm{Re}}

\begin{document}
\widetext


\title{Analytic results for the Anderson Impurity Model}

\author{Vassilis Pandis, Alex C.\ Hewson}
\affiliation{Department of Mathematics, Imperial College London, London SW7 2AZ, United Kingdom}
\date{\today}

\begin{abstract}
In the Renormalised Perturbation Theory (RPT) the Anderson impurity model is interpreted in terms of
renormalised parameters $\mut = (\epsdt, \Deltat, \Ut)$ which are in a one-to-one correspondence with the
bare impurity level $\epsilon_d$, hybridisation $\Delta$ and on-site interaction $U$.  Though the 
renormalised parameters suffice to describe the low-energy fixed point of the model, the parameters themselves usually have to 
be determined using the Numerical Renormalisation Group. In this paper we address this issue by applying the flow equation
method to the particle-hole symmetric model to derive a system of differential equations for $\rmd \mut(U)/ \rmd U$.
To leading order in $\Ut$ this assumes a particularly simple form, permitting its analytic solution in this approximation. 
We find that our results for the renormalised parameters are in good agreement with the NRG and use them
to calculate the Wilson ratio, spin susceptibility and conductance of the impurity. 
\end{abstract}
\maketitle


The Anderson impurity model~\cite{Anderson61} is the canonical model for a magnetic impurity embedded in a non-magnetic
host metal. In its simplest form it comprises an impurity level $\epsilon_d$ and conduction band coupled
through a hybridisation matrix element $\Delta$. To account for the local Coulomb repulsion, an interaction
constant $U>0$ ensures that double occupancy of the impurity level is energetically penalised. The challenge
presented by the model is that for realistic systems $U$ is too large to be treated perturbatively and
alternative techniques have to be devised. Despite its age, the AIM remains a subject of contemporary interest
as a model for the study of transport properties of quantum dots under a bias voltage. These can be fabricated
in a variety of configurations and their tunability renders them very attractive platforms for the study of 
strong correlation physics. 

The AIM and its generalisations have been studied by a variety of methods~\cite{Hewson97},
establishing it as the benchmark model for the development of new methods for strongly correlated systems. In this article we focus on the Renormalised Perturbation Theory (RPT), a standard technique in high-energy
physics that has only relatively recently been applied to condensed matter systems. Amongst other models, it has been applied to the AIM and shown
to be very useful in describing the low-energy features of the single~\cite{Hewson93b,Hewson94, Hewson01,Hewson06,Hewson06b,Hewson06c, Bauer07} and multiple channel models~\cite{Nishikawa10, Nishikawa10b}, as well as its non-equilibrium properties~\cite{Hewson05, Sakano11}.  

Central to the RPT is the concept of a quasi-particle, which is described in terms of two renormalised parameters $\epsdt$ and $\Deltat$ that 
incorporate the one-particle low-energy interactions and thus differ from the bare parameters of the system. The
quasi-particles are weakly interacting through a renormalised interaction $\Ut$, which can be treated perturbatively. 
The failure of ordinary perturbation theory is thus a consequence of an unhelpful division of the
Lagrangian (or Hamiltonian) into non-interacting and interacting components. Through the introduction of 
counter-terms, the RPT achieves a more appropriate arrangement by taking the non-interacting quasi-particles
to constitute the starting point for the expansion, which is to be carried out in the renormalised parameters.

In the low-temperature limit, a number of quantities such as the spin and charge susceptibilities, the
specific heat and the thermopower can be computed exactly given the renormalised parameters. Historically~\cite{Hewson04}, these were determined with the help of the Numerical Renormalisation
Group~\cite{Wilson75, Krishna-murthy80} -- this however restricts the application of RPT to
problems already tractable through the NRG. Recently, a flow equation technique has been developed that allows the renormalised parameters to be determined
exclusively within RPT and without dependence on any external method~\cite{Edwards11, Edwards11b, Edwards13, Pandis14}.
One begins by identifying a variable $\alpha$ of the bare model and a limit $\alpha_0$ in which ordinary perturbation
theory is applicable, and using it to calculate $\mut(\alpha_0) = (\epsdt(\alpha_0), \Deltat(\alpha_0), \Ut(\alpha_0))$; 
note that this limit need not necessarily be physically realisable. The second step is to deduce a system of 
equations to describe $\rmd\mut(\alpha)/\rmd\alpha$ and finally solve it, using knowledge of $\mut(\alpha_0)$ to fix
the integration constants.

In Refs.~\cite{Edwards11, Edwards11b, Edwards13} a magnetic field $h$ was introduced and chosen as the flow variable (a
similar approach in the context of the Functional Renormalisation Group was explored in Ref.~\cite{Streib13}). 
This breaks the $SU(2)$ symmetry of the system and, in the limit $h\rightarrow\infty$, the spin fluctuations are frozen out, rendering it
amenable to a mean-field approach. As $h$ is reduced the interactions are slowly re-introduced, renormalising the 
parameters. Similarly, in Ref.~\cite{Edwards13} the variable $\alpha$ was identified with $\epsilon_d$. By 
starting from $\epsilon_d \rightarrow -\infty$ and gradually reducing the asymmetry factor, renormalised parameters
for the asymmetric model were deduced. Finally in Ref.~\cite{Pandis14} the flow equations in $\Delta$ were studied and
used to calculate renormalised parameters.

In this paper we examine the possibility of using the interaction $U$ as the flow variable. For simplicity we restrict
our discussion to the symmetric model in the absence of a magnetic field, for which $\epsdt=0$, leaving only two renormalised
parameters to be determined. We derive two coupled differential equations to describe the flow of the parameters
with $U$ in terms of derivatives of the renormalised self-energy $\Sigmat(\omega)$. By calculating the leading-order
terms in the derivatives we arrive at a simple system which can be solved in closed form, yielding very simple expressions
for $\mut(U)$. We compare our results to values of the parameters deduced from the NRG and obtain very good agreement
even in the strong correlation regime. From the closed-form expressions we also obtain a very simple expression for the stationary point of $\Ut(U)$.
Finally, we use the parameters to compute the spin susceptibility, Wilson ratio and conductance of the impurity.  

We begin by defining the renormalised quantities $\Deltat = z \Delta$, $\epsdt = z( \epsilon_d + \Sigma(0))$ and $\Sigmat(\omega) = z\Sigma_{r}(\omega)$
where $z = [1 - \Sigma'(0)]^{-1}$ and $\Sigma_{r}(\omega) =  \Sigma(\omega) - \Sigma'(0)\omega - \Sigma(0)$.  The renormalised one-particle irreducible self-energy satisfies 
$\Sigmat(0) = \Sigmat'(0) = 0$, and the renormalised interaction is defined in terms of the bare two-particle reducible four-vertex as $\Ut = z^2 \Gamma_{\uparrow \downarrow}(0,0)$. 
Equivalently, we can set up the RPT by separating the effective Lagrangian of the Anderson model~\cite{Hewson01}
\begin{equation}
\mathcal{L}  = \sum_{\sigma = \uparrow, \downarrow} \overline{d}_\sigma(\tau)  \left(  \partial_\tau - \epsilon_d - i\Delta\right) d_\sigma(\tau) + U n_{\uparrow}(\tau) n_{\downarrow}(\tau), 
\label{eq:lagrangianaim}
\end{equation}
into a renormalised Lagrangian of the same form and a counter-term Lagrangian by writing $\mathcal{L}(\gvec{\mu}) = \tilde{\mathcal{L}}(\mut) + \tilde{\mathcal{L}_{ct}}(\gvec{\lambda}(\mut))$, where
\begin{equation}
\tilde{\mathcal{L}}_{ct}(\gvec{\lambda}) =  \sum_{\sigma = \uparrow, \downarrow}\tilde{\overline{d}}_\sigma(\tau) (\lambda_{2, \sigma} \partial_\tau + \lambda_{1, \sigma})\tilde{d}_\sigma(\tau) + \lambda_3 \tilde{n}_\uparrow(\tau) \tilde{n}_\downarrow(\tau), 
\end{equation}
and the counter-terms are such that $\Sigmat(0) = \Sigmat'(0) = 0$ and $\Ut = \tilde{\Gamma}_{\uparrow \downarrow}(0,0)$.

The renormalised self-energy $\Sigmat(\omega)$ can be computed diagrammatically and the leading term is of order $\Ut^2$~\cite{Hewson93b}.
One finds that 
\begin{equation}
\lambda_2 = - (3-\pi^2/4)\left( \frac{\Ut}{\pi\Deltat}\right)^2 + \mathcal{O}(\Ut^4).
\end{equation}
An important feature of RPT is that the second-order result for $\partial^2_\omega \Sigmat(0)$ is \emph{exact}. 
We now apply the flow equation method of Refs.~\cite{Edwards11b, Edwards13, Pandis14} to the variable $U$.  Note that at particle-hole symmetry $\epsilon_d + U/2=0$, 
$\Sigma(0) = U/2$ and hence $\epsdt=0$; we thus have to deduce only two equations.
 We begin by writing the effective Lagrangian in the RPT formalism for a model with an interaction $U+\delta U$.
\begin{equation}
\mathcal{L}(U + \delta U) = \tilde{\mathcal{L}}(\mut(U + \delta U)) + \tilde{\mathcal{L}_{ct}}(\gvec{\lambda}(\mut(U +\delta U ))).
\label{eq:lagone}
\end{equation}

We can expand around $\mathcal{L}(U)$ by writing
\begin{equation}
\mathcal{L}(U + \delta U) = \mathcal{L}(U) + \delta U \frac{\partial \mathcal{L}}{\partial U} = \mathcal{L}(U) + \mathcal{L}_r (\delta U), 
\label{eq:flowexpansion}
\end{equation}
where the notation $\mathcal{L}_r (\delta U)$ has been introduced for clarity. Note that for the Anderson model the Lagrangian
is linear in the parameters, so Eq.~\eqref{eq:flowexpansion} is \emph{exact} and furthermore $\mathcal{L}_r$ depends \emph{only} on $\delta U$.
We now express $\mathcal{L}(U)$ in terms of renormalised parameters 
\begin{equation}
\mathcal{L}(U + \delta U) = \tilde{\mathcal{L}}(\mut) + \tilde{\mathcal{L}_{ct}}(\gvec{\lambda}(\mut)) + \tilde{\mathcal{L}_r} (\delta U).
\end{equation}
We now use the symbols $\mathcal{L}_0$ and $\mathcal{L}_{ct}$ to denote the first and second terms on the RHS of Eq.~\eqref{eq:lagrangianaim} respectively.
We separate the RHS of Eq.~\eqref{eq:lagone} into an interacting and non-interacting component 
\begin{eqnarray}
\mathcal{L}(U + \delta U) = &\tilde{\mathcal{L}_0}(\mut(U + \delta U)) + \Big[\tilde{\mathcal{L}_I}(\mut(U + \delta U)) + \nonumber \\
&\tilde{\mathcal{L}_{ct}}(\gvec{\lambda}(\mut(U +\delta U )))\Big].
\label{eq:lagtwo}
\end{eqnarray}
Similarly, for Eq.~\eqref{eq:flowexpansion} we have
\begin{eqnarray}
\mathcal{L}(U + \delta U) = \tilde{\mathcal{L}_0}(\mut(U)) + \Big[\tilde{\mathcal{L}_I}(\mut(U)) &+ \tilde{\mathcal{L}_{ct}}(\gvec{\lambda}(\mut(U))) \nonumber \\
&+ \mathcal{L}_r (\delta U)\Big].
\label{eq:lagthree}
\end{eqnarray}
In Eqs.~\eqref{eq:lagtwo}, \eqref{eq:lagthree} we have two different ways of rewriting the same bare Lagrangian. In Eq.~\eqref{eq:lagtwo}
the Lagrangian $\tilde{\mathcal{L}_0}(\Ut(U + \delta U))$ describes the non-interacting quasi-particles which
interact through the term in square brackets (the perturbation) and which gives rise to the renormalised self-energy $\Sigmat(\omega; \mut(U + \delta U))$. By contrast,
in Eq.~\eqref{eq:lagthree} the free quasi-particles are described in terms of $\mut(U)$. This results in an additional interaction term, $\mathcal{L}_r (U)$
which generates a corresponding interaction vertex. This must be appropriately included in the diagrammatic expansions which now give rise to a self-energy
$\Sigmat^{(2)}(\omega; \mut(U), \delta U)$. In the trivial case of $\delta U=0$ this does not contribute at all and we see that
$\Sigmat^{(2)}(\omega; \mut(U), \delta U=0)=\Sigmat(\omega; \mut(U + \delta U))$.

To relate the two self-energies we express the bare propagator in two different ways:
\begin{align}
G(\omega) &= \frac{z(U + \delta U)}{\omega + i \Deltat(U+\delta U) - \Sigmat(\omega; \mut(U + \delta U))} \nonumber \\
                 &= \frac{z(U)}{\omega + i \Deltat(U) -\Sigmat^{(2)}(\omega; \mut(U), \delta U)}.
\label{eq:twogreens}
\end{align}
By taking the reciprocal of Eq.~\eqref{eq:twogreens}, equating the derivatives at zero and taking the real part we find
\begin{equation}
z(U; \delta U) = \left[ 1 -\derivat{\frac{\partial\Re \Sigmat^{(2)}(\omega; \mut(U), \delta U)}{\partial \omega}}{\omega=0}\right]^{-1}.
\label{eq:zfrac}
\end{equation}
By expanding the bracket of Eq.~\eqref{eq:zfrac} in $\delta U$ and retaining the leading term we find that
\begin{equation}
z(U; \delta U) = 1 + q(\mut(U))\delta U,
\label{eq:zbar}
\end{equation}
where 
\begin{equation}
q(\mut(U)) = \frac{\partial}{\partial U} \derivat{\frac{\partial\Re \Sigmat^{(2)}(\omega; \mut(U), \delta U)}{\partial \omega}}{\omega=0}.
\end{equation}
From the definition of $z(U; \delta U)$ and Eq.~\eqref{eq:zbar} we find that
\begin{equation}
\frac{\partial \ln \Deltat(U)}{\partial U} = q(\mut(U)).
\label{eq:flowdelta}
\end{equation}
This constitutes the flow equation for $\Deltat(U)$. To determine $q(\mut(U))$ we appeal to the second-order calculation. We can
account for $\mathcal{L}_r$ simply by substituting $\Ut \rightarrow \Ut + z \delta U$; this renders the cancellation with $\lambda_2$ incomplete, yielding 
\begin{equation}
q(U) = \frac{\xi \Ut(U)}{\Deltat(U)}, 
\end{equation}
where we have introduced a constant  
\begin{equation}
\xi = - (6-\pi^2/2)  \frac{1}{\pi^2\Delta}<0.
\end{equation}

To derive the flow equation for $\Ut$ we differentiate Eq.~\eqref{eq:twogreens} twice to find
\begin{equation}
{\partial^2_\omega}\Sigmat(\omega; \mut(U + \delta U)) = z(U; \delta U)\partial^2_\omega \Sigmat^{(2)}(\omega; \mut(U), \delta U)).
\label{eq:secondderivatives}
\end{equation}
The LHS can be deduced exactly from the second-order calculation and will be proportional to $\Ut^2/\Deltat^3$. We can use the same expression
to calculate the RHS of Eq.~\eqref{eq:secondderivatives}. Note that due to the presence of the infinitesimal vertex this will not be exact and
will receive higher-order corrections, which in this paper we do not attempt to calculate. We arrive at the following equation
\begin{equation}
\frac{ \Ut^2 (U + \delta U)}{\Deltat^3 (U + \delta U)} = z(U; \delta U)\frac{(\Ut(U) + z \delta U)^2}{\Deltat^3(U)}.
\label{eq:Uflow1}
\end{equation}
We can now combine Eq.~\eqref{eq:Uflow1} and Eq.~\eqref{eq:flowdelta} to derive the complete system of differential flow equations
\begin{align}
\frac{\partial \Deltat}{\partial U} &= \xi \Ut \nonumber \\
\frac{\partial \Ut}{\partial U} &= 2 \xi \frac{\Ut^2}{\Deltat} + \frac{\Deltat}{\Delta} .
\label{eq:system}
\end{align}


We proceed to solve Eq.~\eqref{eq:system} by eliminating $\Ut$ to find
\begin{equation}
\frac{d^2 \Deltat}{d U^2} = \frac{2}{\Deltat} \left( \frac{d\Deltat}{dU}\right)^2 - \kappa^2 \Deltat, 
\end{equation}
where $\kappa = \sqrt{|\xi|/\Delta}$. We solve this, fixing the constants of integration 
by requiring that $\Deltat(0) = \Delta$ and that
$\Deltat(U)$ be quadratic in $U$ for $U\rightarrow0$,  so as to be compatible with the results of ordinary perturbation theory~\cite{Yosida70, Yamada75, Yosida75}. The result is
\begin{align}
\Deltat(U) &= \Delta \sech(\kappa U), \label{eq:deq}\\
\Ut(U) &= \frac{\sinh(\kappa U)}{\kappa \cosh^2(\kappa U)} \label{eq:uteq}.
\end{align}

\begin{figure}
\centering
\includegraphics[scale=0.29]{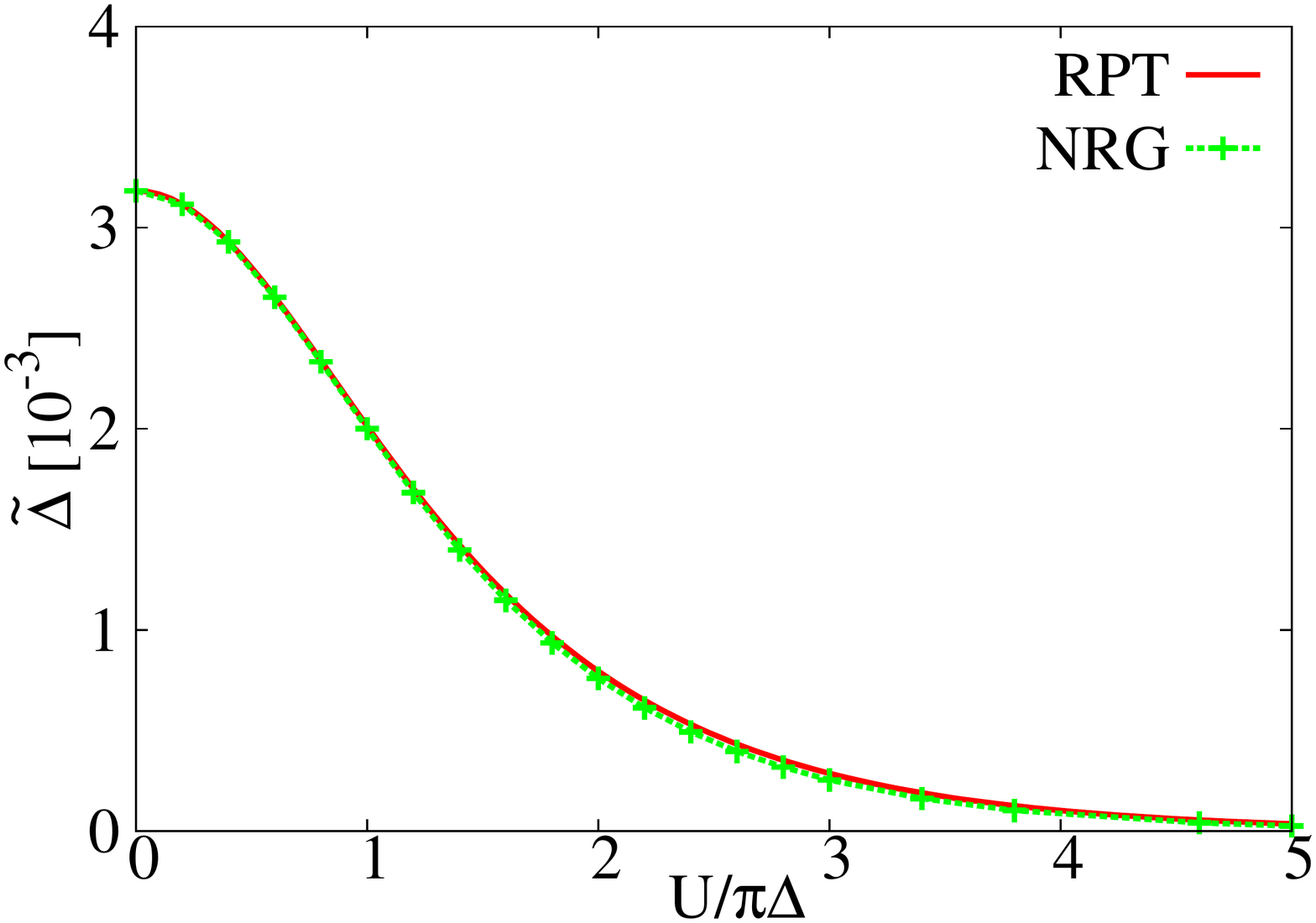}
\caption{The parameter $\Deltat(U)$ from Eq.~\eqref{eq:deq} compared to the NRG results.}
\label{fig:deltaflow}
\end{figure}

\begin{figure}
\centering
\includegraphics[scale=0.29]{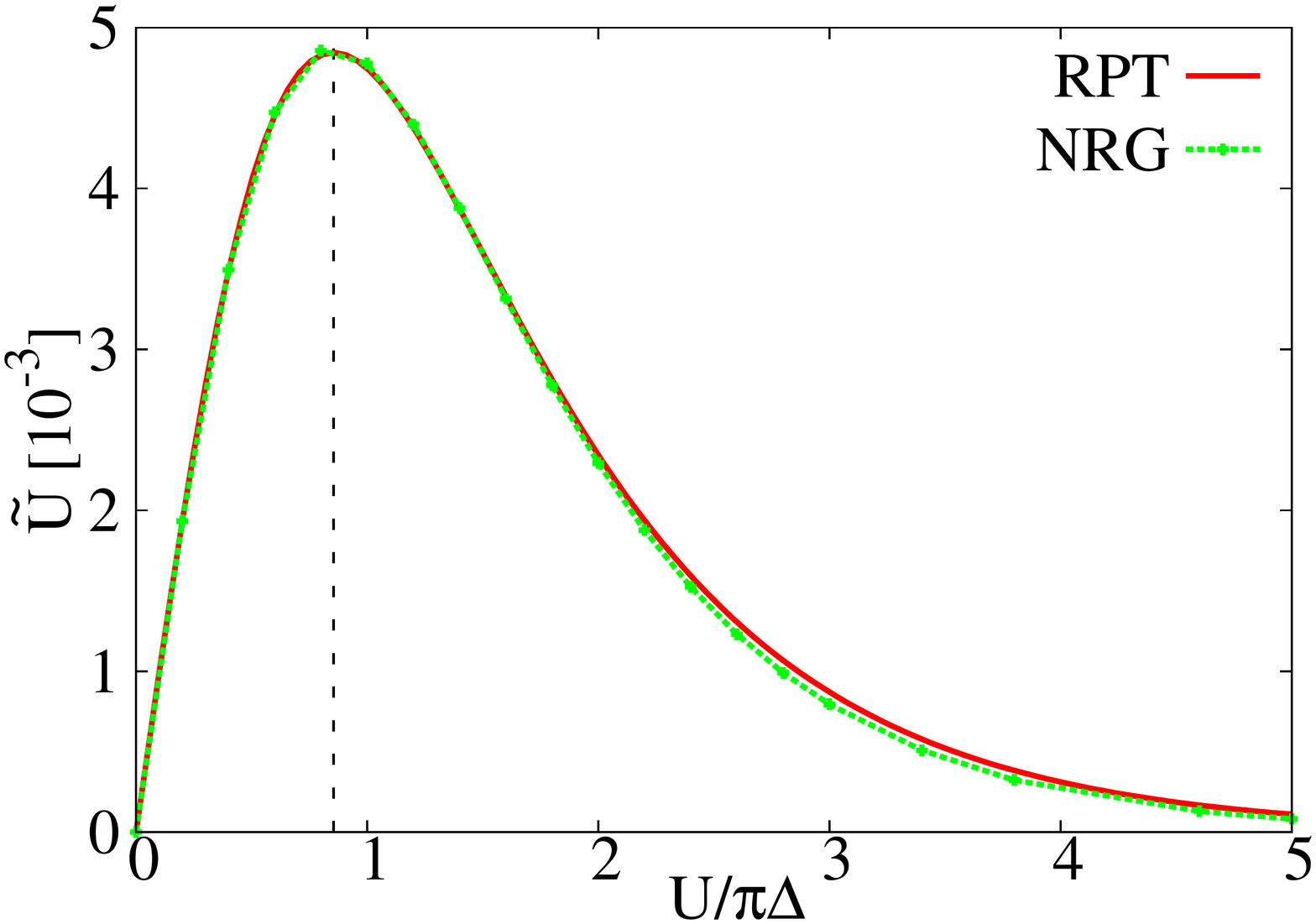}
\caption{The parameter $\Ut(U)$ from Eq.~\eqref{eq:uteq}, compared to the NRG results.}
\label{fig:uflow}
\end{figure}

\begin{figure}
\centering
\includegraphics[scale=0.29]{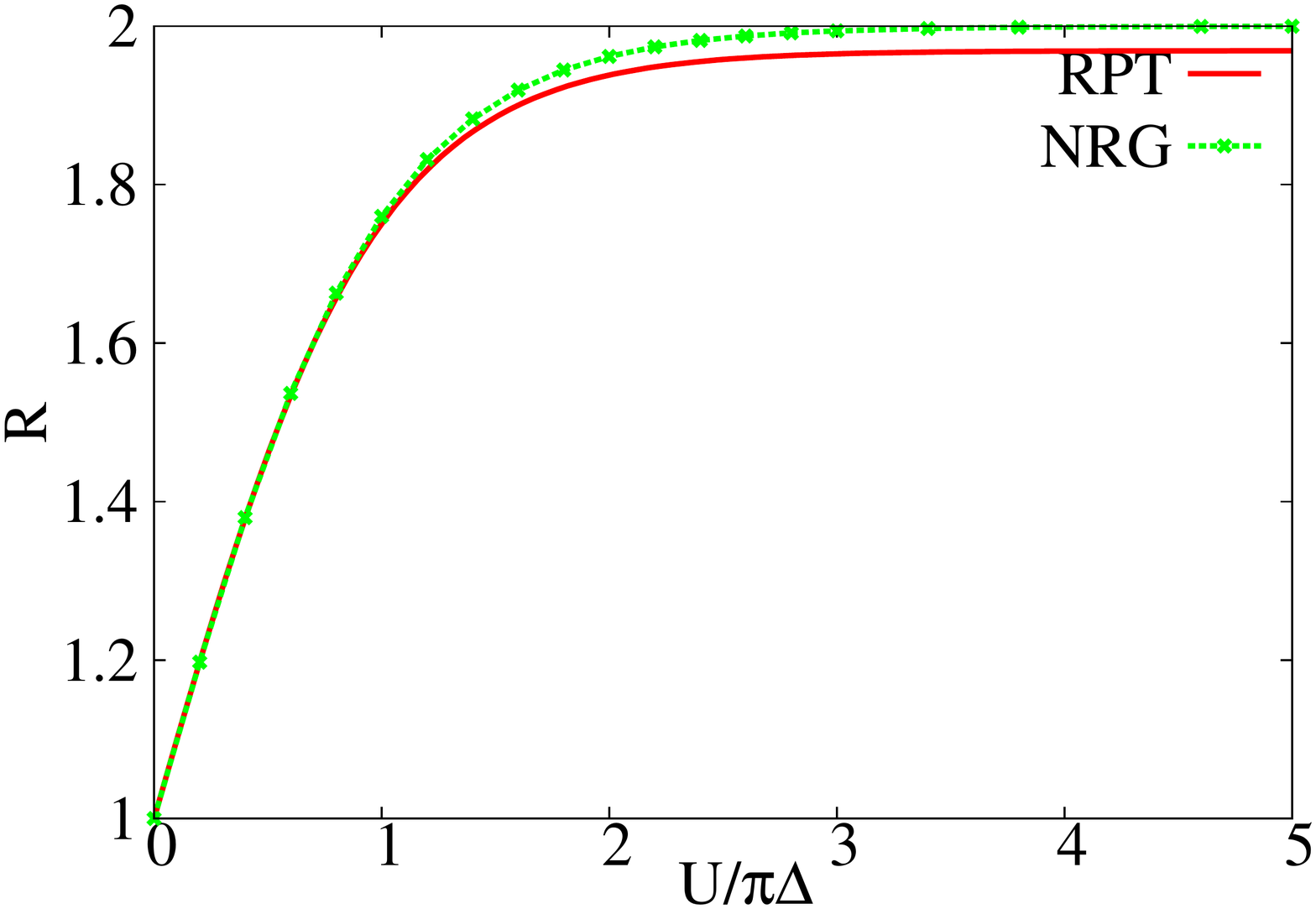}
\caption{The Wilson ratio $R = 1+ \Ut/\pi\Deltat$.}
\label{fig:flowratio}
\end{figure}

These equations are compared to data from the NRG in Fig.~\ref{fig:uflow}. We find
that at small values of $U/\pi\Delta$ both renormalised parameters are in excellent agreement with the NRG.
From the plots we also confirm that the maximum of $\Ut(U)$ is faithfully reproduced by Eq.~\eqref{eq:uteq}, 
which predicts that the stationary point of $\Ut(U)$ occurs at
\begin{equation}
\frac{U_s}{\pi\Delta} = \frac{\textrm{tanh}^{-1}\left(\frac{1}{\sqrt{2}}\right)}{\sqrt{6 - \pi^2/2}}\approx 0.85397,
\end{equation}
and is indicated by the vertical line in Fig.~\ref{fig:uflow}. We note that this is only $0.25\%$ smaller than the $U_s/\pi\Delta$ that can be deduced from the Bethe Ansatz results of Ref.~\cite{Andrei83, Wiegmann83}.
As $U$ increases small
discrepancies arise and become more severe with $U$, with the RPT estimates of the parameters being larger than the NRG values. 
We attribute this to the fact that $\Ut/\pi\Deltat$ is approaching 1 so higher-order terms in $q(U)$ and the RHS of 
Eq.~\eqref{eq:secondderivatives} start to contribute. The Wilson ratio $R = 1+ \Ut/\pi\Deltat$ is plotted alongside data from the NRG in Fig.~\ref{fig:flowratio}.
In the Kondo limit we obtain that $R \rightarrow 1+ (6-\pi^2/2)^{-1/2}  \approx 1.9689$, 
which is close to $2$, the exact result.

\begin{figure}
\centering
\includegraphics[scale=0.29]{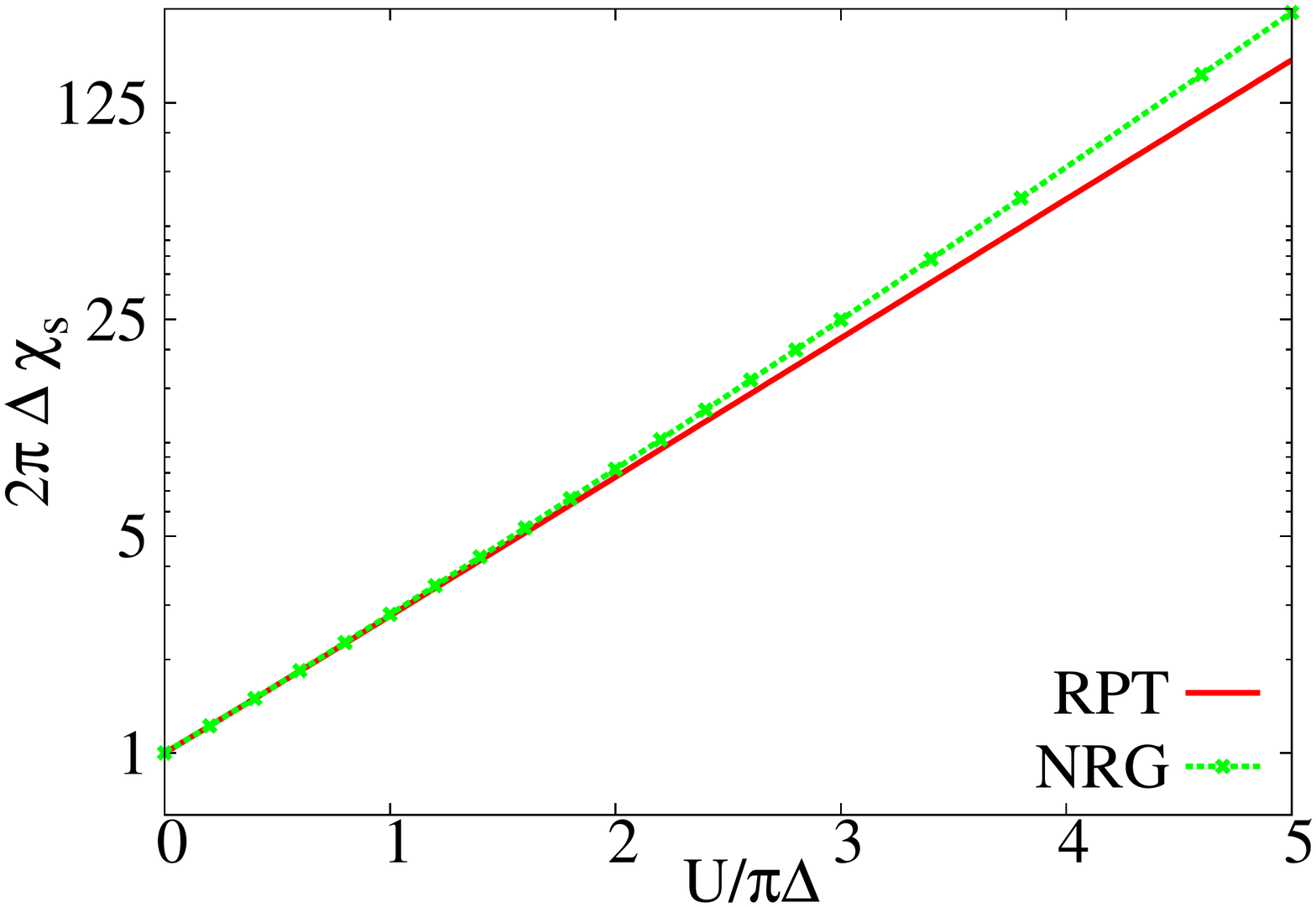}
\caption{RPT results for the reduced spin susceptibility $\overline{\chi}_s = 2\pi\chi_s$, shown as a  function of $U/\pi\Delta$.}
\label{fig:xs}
\end{figure}

Having computed the renormalised parameters we can now proceed to calculate physical properties of the model. In the
RPT the reduced spin susceptibility $\overline{\chi}_s = 2\pi\Delta\chi_s$ is given at particle-hole symmetry by~\cite{Hewson93b} 
\begin{equation}
\overline{\chi}_s = \frac{\Delta}{\Deltat} \left( 1 + \frac{\Ut}{\pi\Deltat}\right).
\end{equation}
Note that this expression is \emph{exact}, provided of course that $\Deltat$ and $\Ut$ are exactly known. From the results in
in Fig.~\ref{fig:xs} we see that the RPT results are in excellent agreement with the NRG at weak coupling, though this progressively
deteriorates as $U/\pi\Delta$ increases. In particular, the RPT result for $\ln\overline{\chi}_s$ displays a greater convexity than the curve from
the NRG and thus somewhat overestimates $\ln\overline{\chi}_s$ at large $U$.

\begin{figure}
\centering
\includegraphics[scale=0.29]{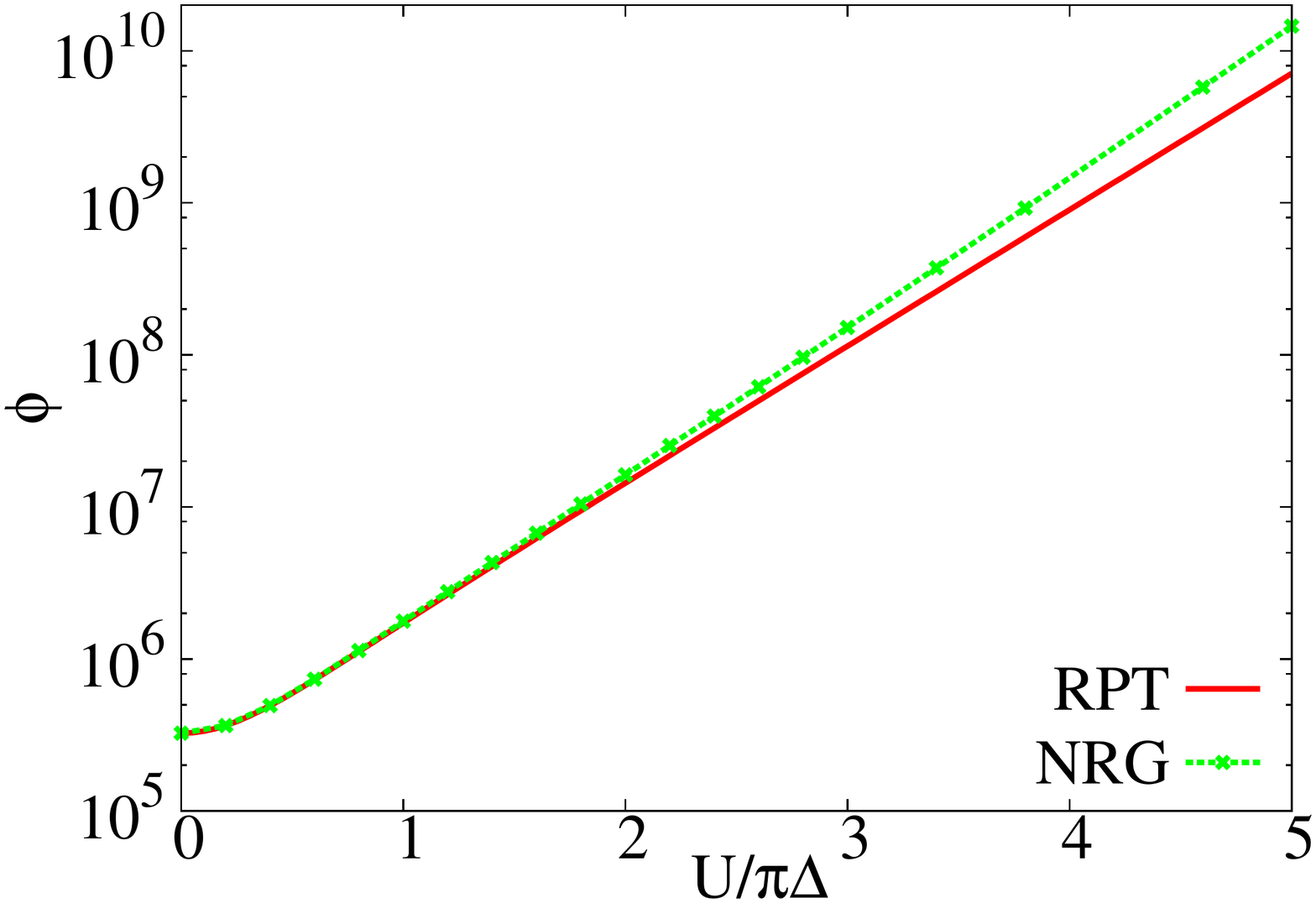}
\caption{The $\phi$ coefficient of the conductance, defined through $\sigma(T)/\sigma(0) = 1 + (k_B T)^2 \phi$. }
\label{fig:sigma}
\end{figure}

Finally, we turn our attention to the impurity conductivity, which can be written~\cite{Hewson93b} as $\sigma(T)/\sigma(0) = 1 + (k_B T)^2 \phi$,
where
\begin{equation}
\phi = \frac{\pi^2}{3\Deltat^2}\left[1 + 2(R-1)^2\right].
\end{equation}
The quantity $\phi$ is  plotted in Fig.~\ref{fig:sigma}, where we see that it is in excellent agreement with the NRG at small and intermediate couplings,
and it is only at very strong interactions that significant discrepancies can be observed. We attribute these to the fact that our
method does not fully capture the exponential suppression of $\Deltat$ at very large $U$. 

In summary, by starting from the limit $U\rightarrow0$ and slowly increasing $U$,  we derived simple approximate 
closed form expressions for the renormalised parameters of the symmetric model. We found these to be in very good agreement
with the NRG, even when $U$ is large, but with discrepancies arising in the $U\rightarrow\infty$ limit. Using our expressions
we calculated the Wilson ratio, spin susceptibility and impurity conductance, and found them to be in good agreement with the NRG. 
We remark that the inclusion of the next-to-leading terms in $\Ut$ or a ladder-based resummation is likely to improve 
the accuracy of our approach in the large-$U$ limit and that our method should in principle be 
easy to generalise to the asymmetric model.

V.P.\ acknowledges the support of the Engineering and Physical Sciences Research Council.
\bibliographystyle{apsrev4-1}
\bibliography{bibliography}

\end{document}